\documentclass[aps,prx,twocolumn,superscriptaddress]{revtex4-1}

\usepackage{tabularx}
\usepackage{graphicx}
\usepackage{dcolumn}
\usepackage{amsfonts,amsmath,amssymb,bm}
\usepackage{xcolor}
\usepackage[utf8]{inputenc}
\usepackage[colorlinks=true,allcolors=blue]{hyperref}
\usepackage{multirow}
\begin{document}
\def \CS {Cr$_3$Si}	

\title{Impact of Ge, Ga, and Al doping on the mechanical and electronic properties of \CS: insights from first-principles calculations}

\date{\today}
\author{Siavash Karbasizadeh}
\email[Corresponding author: ]{s\_karbasizadeh@ucsb.edu}
\affiliation{Materials Department, University of California, Santa Barbara, California 93106-5050, USA}

\author{Mohammad Amirabbasi}
\affiliation{Technical University of Darmstadt, Materials Modelling Division, Otto-Berndt-Straße 3, Darmstadt D-64287, Germany}

\begin{abstract}
This study systematically investigates the effects of Ge, Ga, and Al doping on the mechanical and electronic properties of cubic \CS~using first-principles density functional theory (DFT). Doping increases lattice constants from 4.50 Å for undoped \CS~to 4.51–4.53 Å (Ge), 4.52–4.54 Å (Ga), and 4.51–4.54 Å (Al) as doping concentrations increase from 12.5 $\%$ to 50 $\%$. Negative formation enthalpies across all configurations confirm thermodynamic stability, with values ranging from -0.35 eV/atom for undoped \CS~to -0.33 eV/atom (Ge), -0.31 eV/atom (Al), and -0.25 eV/atom (Ga) at 50 $\%$ doping. Mechanical properties exhibit significant degradation with increased doping: bulk modulus decreases from 248.7 GPa for undoped \CS~to 241 GPa (12.5 $\%$), 238 GPa (25 $\%$), 235 GPa (37.5 $\%$), and 231 GPa (50 $\%$) for Ge doping, with similar trends for Ga (230 GPa at 50 $\%$) and Al (232 GPa at 50 $\%$). Shear modulus and Young’s modulus follow similar reductions, with shear modulus going from 158.9 GPa to 147 GPa (Ge), 145 GPa (Ga), and 147 GPa (Al) at 50 $\%$ doping. Elastic anisotropy increases notably with Al and Ga doping, while Ge maintains a relatively isotropic behavior. The wave velocities and Debye temperatures decrease for all dopants, with Debye temperature dropping from 720 K for undoped \CS~to 700 K (Ge), 685 K (Ga), and 690 K (Al) at 50 $\%$ doping, reflecting a softer lattice and diminished thermal conductivity. While Al and Ga doping introduce higher anisotropy and reduce mechanical rigidity, Ge doping preserves isotropic mechanical behavior, making it the most suitable dopant for applications requiring balanced mechanical and thermal properties. These findings offer critical insights into tailoring \CS~-based alloys for high-performance applications, highlighting trade-offs between stiffness, anisotropy, and thermal performance.
\end{abstract}

\maketitle

\section{INTRODUCTION}
The Cr-Si binary silicides have received great attention due to their high thermal stability and excellent mechanical properties \cite{C9CY00533A}. Owing to their better oxidation resistance over their silicide counterparts, their usage in the industry has been extensive \cite{doi:10.1021/ic5024482}. 
Of the binary phases of Cr-Si, the cubic \CS~phase stands out~\cite{IOROI2024102690,Wijin1991,PhysRevB.25.5856,Mihailov1979,Anton1990,RAJ1995583,Soleimani-Dorcheh2015,https://doi.org/10.1002/maco.201307423}. 
It has been reported to be stable in a temperature range down to 6 K \cite{Wijin1991,PhysRevB.25.5856} with a Pauli paramagnetic behavior in the temperature range between 4.2 and 300 K \cite{Mihailov1979}.
\CS~provides a high melting point (T$_m$ $>$ 1700 $^\circ$C), high temperature strength, and high creep resistance \cite{Anton1990,RAJ1995583}. Recent investigations have also shown this structure to have a promising oxidation and nitridation resistance at ultra-high temperatures (T $>$ 1200 $^\circ$C) \cite{Soleimani-Dorcheh2015,https://doi.org/10.1002/maco.201307423}. This intermetallic system, however, does not meet all the criteria for high-temperature structural applications.
Of the requirements lacking in this material are high microstructural thermal stability, and room-temperature ductility and toughness \cite{SHAH1992402,GALI20093823}.

\begin{figure}
\includegraphics*[scale=0.08]{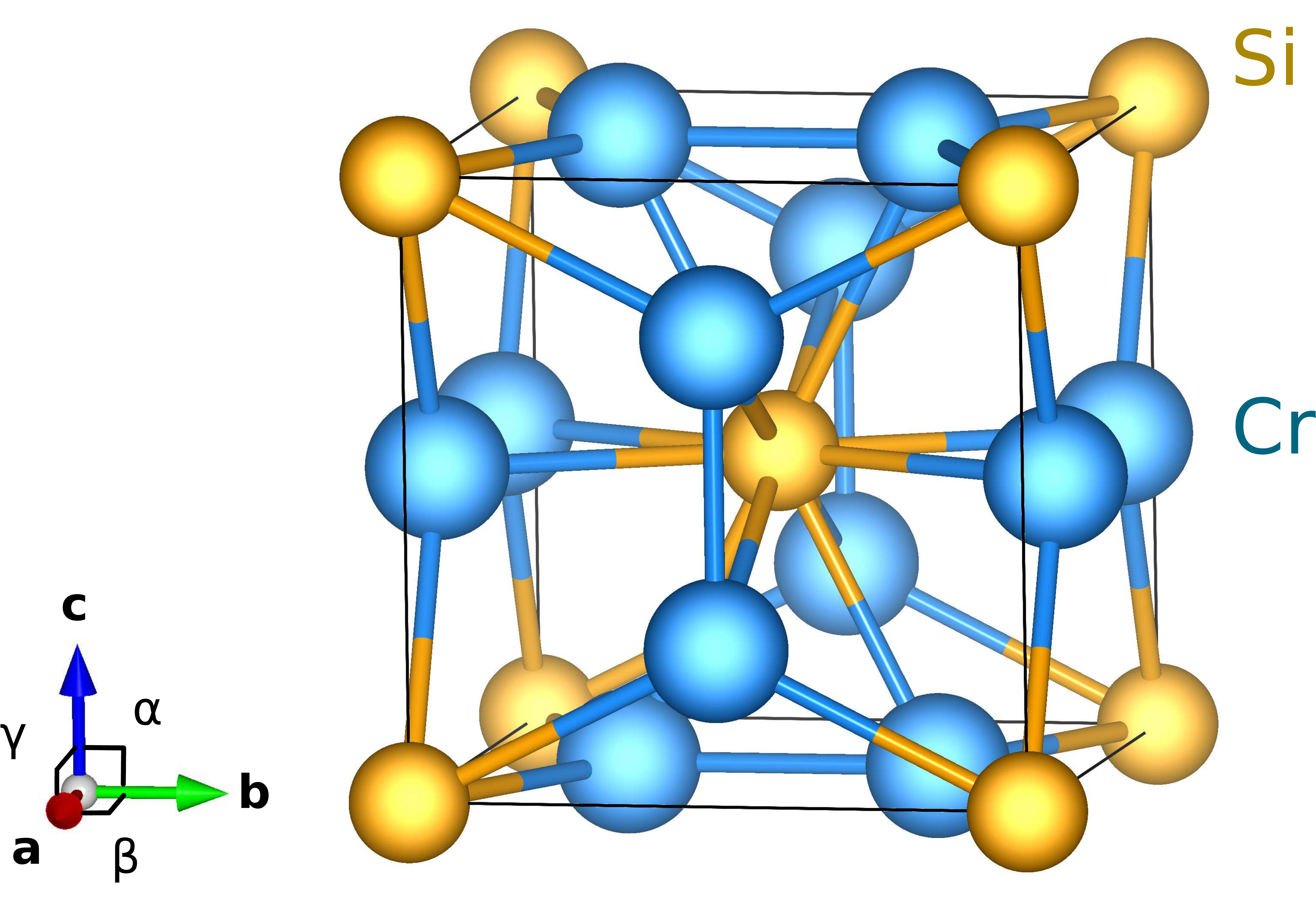}
\caption{
The 8-atom conventional unitcell of \CS. Silicon atoms are shown in yellow, while chromium atoms are shown in blue, as indicated. 
}
\label{Fig1}
\end{figure}

According to geometric point, this cubic structure belongs to an A15 cubic structure with a space group of $Pm\bar{3}n$ in which Cr and Si occupy 6c (0.25, 0, 0.5) and 2a (0, 0, 0) of the Wyckoff positions, respectively (Fig.~\ref{Fig1}). One of the prime approaches to modify and tune the efficiency of this system is alloying
which has seen some interesting improvement of characteristics in the structure to alleviate the deficiencies. Soleimani-Dorcheh \textit{et al.} \cite{Soleimani-Dorcheh2015} showed through measurements that incorporating Ge into the material in small concentrations, substituting Si, can reduce the oxidation and evaporation rates. Raj \cite{RAJ1995229,RAJ1995583} concluded that an alloy of Mo and \CS~can be utilized as erosion and oxidation resistant coatings on turbine airfoils. 
Cruse \textit{et al.} \cite{CRUSE1997410} described improvements of toughness in \CS~alloyed with Mo, with possibilities of improved high temperature properties.
Raj \textit{et al.} \cite{RAJ1999743} also reported improvements in creep properties of \CS~when alloyed with Mo. 
Mo in all cases sits in the place of Cr. Gali \textit{et al.}  \cite{GALI20093823} showed that addition of very small levels of Re and Ce can slow down the coarsening rate at high temperatures. Both elements substitute Cr in the structure.  

Looking at the improvements made by doping the material, a first-principles study on
the mechanical properties of cubic \CS~alloys can prove informative in future endeavors. This study aims to systematically investigate the effect of Ge, Ga, and Al doping on the mechanical properties of \CS~using first-principles calculations.
Ge was incorporated into the material in one previous study \cite{Soleimani-Dorcheh2015} and a theoretical look at the mechanical behavior of the alloy would provide better understanding. Al is chosen as it has shown to have significant effects on the decrease of electrical resistivity and increase of thermoelectric efficiency in high temperatures when alloyed in small concentrations with CrSi$_2$ binary silicide \cite{PAN2007245}. Given the promising results of Ge and Al doping in binary silicides, Ga could potentially portray similar properties. Al and Ge have shown preference to substitute on Si sites. All three elements are therefore substituted exclusively with Si in our set of calculations. 
The doping of \CS~with Ge, Ga, and Al appears to be experimentally feasible, as supported by both theoretical insights and existing literature on related silicide systems. These precedents suggest that doping \CS~with these elements should not pose significant experimental challenges, although precise synthesis methods and conditions (such as controlling dopant concentration and ensuring homogeneity) would need to be carefully optimized.
By analyzing changes in elastic moduli, wave velocities, Debye temperature, and anisotropy indices, we seek to provide a detailed understanding of the structural stability and mechanical behavior of these doped alloys. The insights gained from this research will guide the design of \CS-based materials with tailored properties for specific high-performance applications.
This article is organized as follows. The details of the calculations are described in Sec. \ref{II}, while the results and needed discussion is given in Sec. \ref{III}. The article is then concluded in Sec. \ref{IV}.

%%%%%%%%%%%%%%%%%%%%%%%%%%%%%%%%%%%%%%%%%%%%%%%%%%%%%%%%%%%%%%%%
%%%%%%%%%%%%%%%%%%%%%%%%%%%%%%%%%%%%%%%%%%%%%%%%%%%%%%%%%%%%%%
%%%%%%%%%%%%%%%%%%%%%%%%%%%%%%%%%%%%%%%%%%%%%%%%%%%%%%%%%%%%%%%%%%%%%%%%%%%%%%%%%%%%%%%%%%%%%%%%%%%%%%%%%%%%%%%%
\begin{figure*}
	\includegraphics*[scale=0.68]{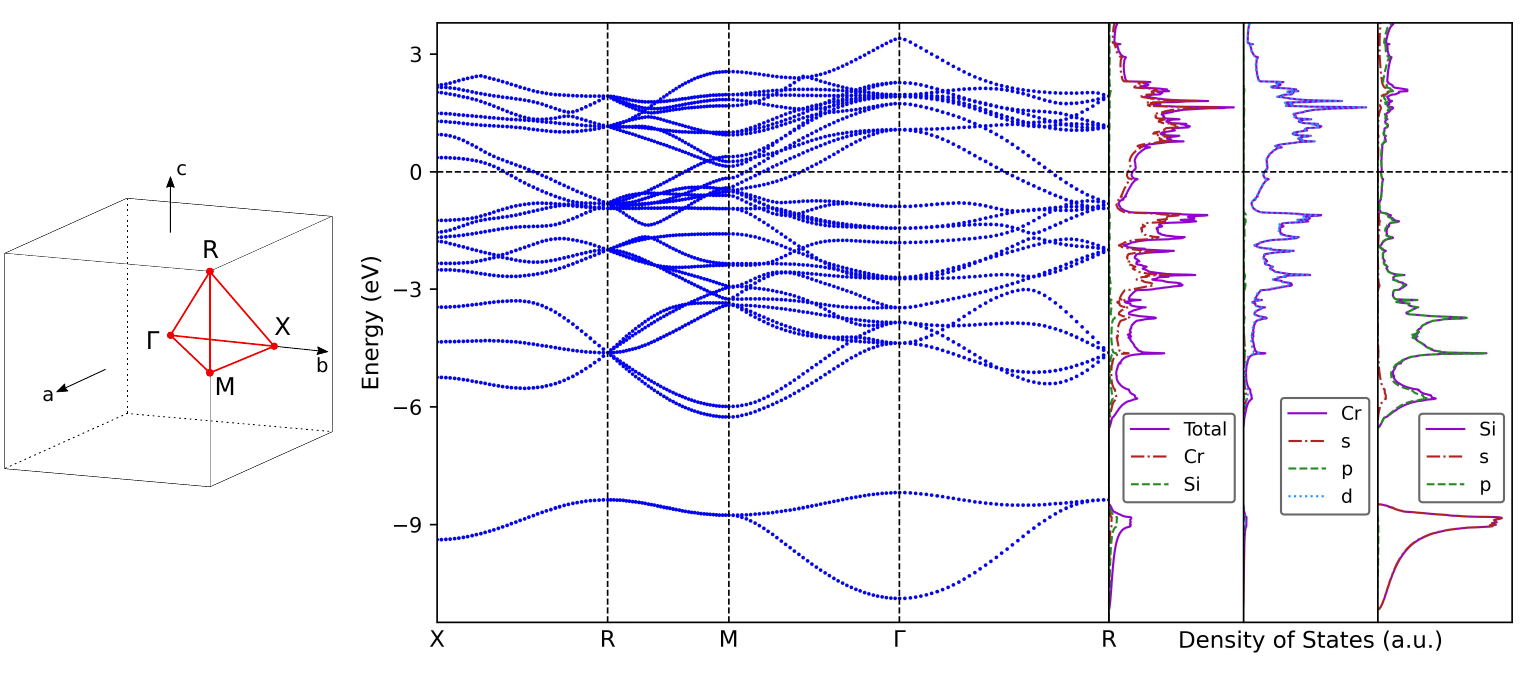}
	\caption{
		Band structure of \CS~alongside densities of state. The specific atomic orbitals contributing to the orbital-projected densities of states are indicated. The path of high-symmetry points considered is shown for clarity. 
	}
	\label{Fig2}
\end{figure*}
%%%%%%%%%%%%%%%%%%%%%%%%%%%%%%%%%%%%%%%%%%%%%%%%%%%%%%%%%%%%%%%%%%%%%%%%%%%%%%%%%%%%%%%%%%%%%%%%%%%%%%%%%%%%%%
%%%%%%%%%%%%%%%%%%%%%%%%%%%%%%%%%%%%%%%%%%%%%%%%%%%%%%%%%%%%%%%%%%%%%%%%%%%%%%%%%%%%
\begin{figure*}
\includegraphics*[scale=0.58]{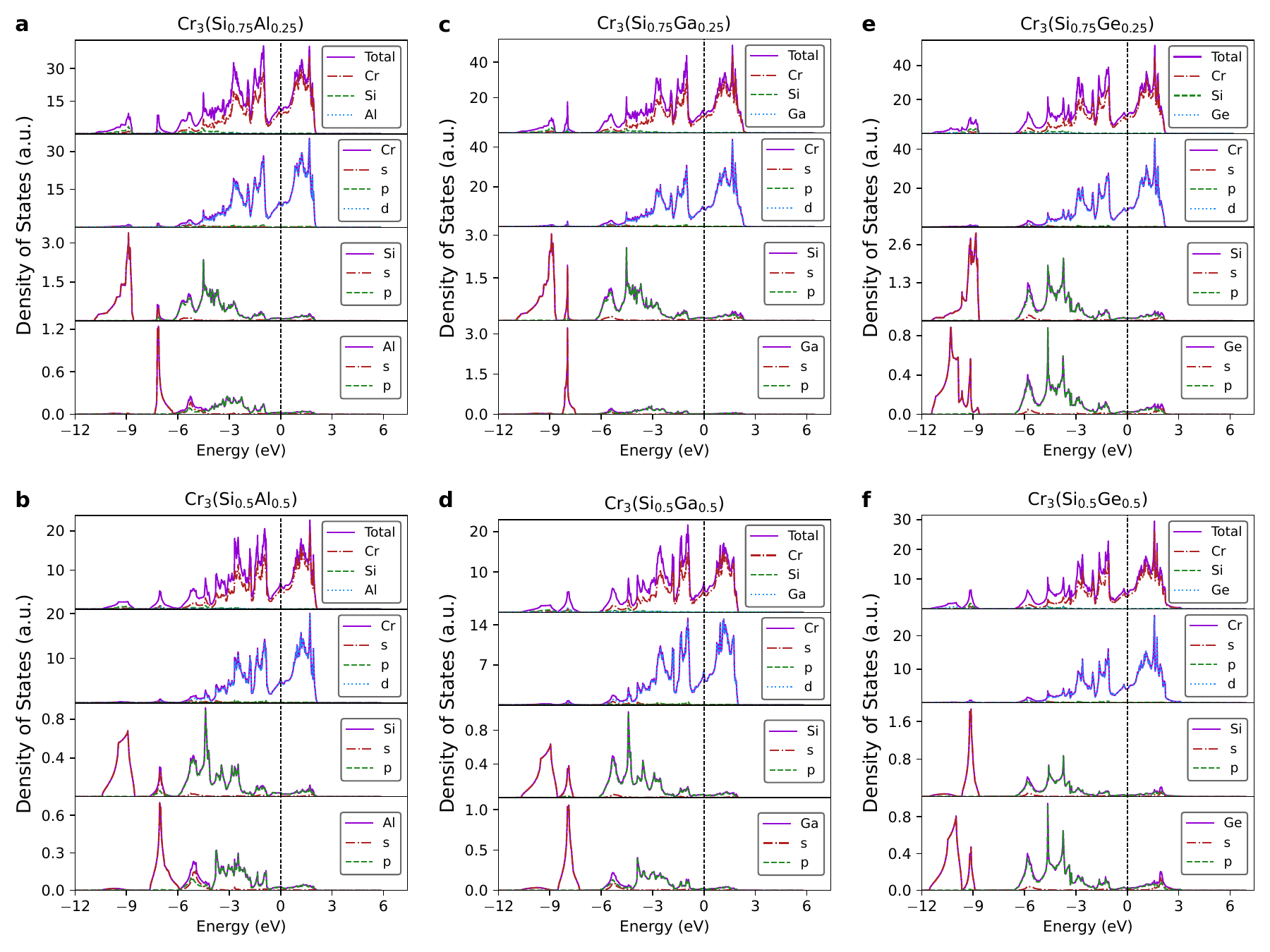}
\caption{
The DOS of doped \CS~ at alloy concentrations of 25\% and 50\%. The DOS plots reveal the contributions of dopants (Ge, Ga, and Al) to the electronic structure.
}
\label{Fig9}
\end{figure*}
%%%%%%%%%%%%%%%%%%%%
%%%%%
\begin{table}[!htp]
\caption{
Calculated and experimental~\cite{CUI2017887} lattice parameters for the 8-atom conventional cell of~\CS.
}
\setlength{\tabcolsep}{10pt}
\begin{tabular}{cccc}
\hline
\hline
& Calc. (GGA) & Calc. (HSE) & Exp.  \\
\hline
$a$ (\AA{})  & 4.50 & 4.43 & 4.57 \\ 
\hline
\hline
\end{tabular}
\label{T0}
\end{table}
%%%%%%%%%%%%%%
%%%%%%%%%%%%%%%%%%%%%%%%%%%%%%%%%%%%%%%%%%%%%%%%%%%%%%%%%%%%%
\begin{table*}[htb!]
\caption{
Elastic constants of \CS~shown alongside prior calculated and experimental values. 
}
\setlength{\tabcolsep}{3pt}
\begin{tabular*}{\textwidth}{c @{\extracolsep{\fill}} cccc}
\hline
\hline
Method & Ref. & $C_\mathrm{11}$(GPa) & $C_\mathrm{12}$(GPa) & $C_\mathrm{44}$(GPa) \\
\hline
VASP-PBE   & This work         & 507.6 & 138.1 & 150.7\\ 
CASTEP-PBE & \cite{Ren_2018}   & 480.8 & 136.1 & 149.4\\
CASTEP-PBE & \cite{wei2018}    & 490.2 & 127.9 & 145.6\\
Exp.       & \cite{BEI2004875} & 416.0 & 98.7  & 129.0\\
\hline
\hline
\end{tabular*}
\label{T1}
\end{table*}
%%%%%%%%%%%%%%%%%%%%%%%%%%%%%%%%%%%%%%%%%%%%%%%%%%%%%%%%%%%%%%%%%%%%%%5
\section{COMPUTATIONAL METHODOLOGY}\label{II}
Density functional theory (DFT) \cite{PhysRev.140.A1133,PhysRev.136.B864} calculations using the projector augmented wave (PAW) \cite{PhysRevB.50.17953,PhysRevB.59.1758} method is employed as implemented in the Vienna ab initio simulation package (VASP) \cite{KRESSE199615}. The exchange correlation energy is described through the generalized gradient approximation (GGA) of Perdew-Burke-Ernzerhof (PBE) \cite{PhysRevLett.77.3865}. Hybrid functional of Heyd-Scuseria-Ernzerhof (HSE) \cite{10.1063/1.1564060} is also used for preliminary tests on the unitcell, with the standard mixing parameter of $\alpha=0.25$. The plane wave basis is used with the 600 eV energy cut-off and the Hellmann-Feynman forces on each atom are converged to 0.001 eV/\AA{}. The smearing method of Methfessel-Paxton is used in the calculations and smearing width of 0.2 eV is selected. Self-consistent calculations are considered to be converged when the energy convergence is less than 10$^{-5}$ eV. Valence electrons of Cr 3$s^2$3$p^6$3$d^5$4$s^1$, Si 3$s^2$3$p^2$, Ge 3$d^{10}$4$s^2$4$p^2$, Ga 3$d^{10}$4$s^2$4$p^2$, and Al 3$s^2$3$p^1$ are adopted in PAW pseudopotentials. The first Brillouin zone is sampled by the $\Gamma$-centered k-point mesh of 12$\times$12$\times$12 in accordance with the Monkhorst-Pack scheme \cite{PhysRevB.13.5188} for the conventional unitcell of \CS~shown in Fig. \ref{Fig1}. The atomic configurations for the study of the alloys at different compositions are generated using the Integrated Cluster Expansion Toolkit (ICET) \cite{https://doi.org/10.1002/adts.201900015}. To study the alloying effects at concentrations 12.5, 25, 37.5, and 50 at \%, we used $\sqrt{2}\times\sqrt{2}\times\sqrt{3}$, 1$\times$1$\times$2, $\sqrt{6}\times\sqrt{2}\times\sqrt{2}$, and 1$\times$1$\times$1 supercells based on the conventional unitcell. Scaled down k-meshes are used for the corresponding supercells. 
In our study, the doping was considered in an ordered manner, where specific substitution sites were chosen for the dopants (Ge, Ga, and Al) within the \CS~lattice. We opted for this approach to maintain a well-defined comparison of how each dopant affects the mechanical properties and anisotropy. By selecting an ordered configuration, we could systematically evaluate the changes introduced by the dopants, ensuring consistency in our calculations.
The formulas used for calculating the mechanical properties are given in Appendix \ref{A1}.
%%%%%%%%%%%%%%%%%%%%%%%%%%%%%%%%%%%%%%%%%%%%%%%%%%%%%%%%%%
\section{RESULTS AND DISCUSSIONS}\label{III} 
\subsection{Electronic and mechanical properties of pristine \CS} 
Here, we discuss the electronic properties of pure \CS. We start from the experimental CIF~\cite{CUI2017887} file and optimize the ionic positions as well as lattice vectors through DFT calculations. To this end, we consider the conventional unitcell which includes 8 ions as illustrated in Figure~\ref{Fig1}. Table \ref{T0} summarizes the experimental and DFT lattice parameters of pristine \CS. The obtained results show that system preserves its cubic symmetry after geometric optimization. In DFT, the accurate description of electron-electron correlation is contingent upon the choice of exchange-correlation functional~\cite{science-2008, PhysRevLett.77.3865}. To determine the most suitable functional for further calculations, it is essential to compare the DFT results with corresponding experimental values. According to Table \ref{T0}, based on the comparison between the obtained lattice constants and experimental values, GGA calculations provide more accurate results than the HSE approach. Therefore, we proceed with deriving the material properties using GGA functional.

The calculated band structure of \CS~along the X-R-M-$\Gamma$-R high-symmetry path is shown in Fig. \ref{Fig2}. The energy bands are plotted relative to the Fermi level ($E_{\mathrm{F}}$), set to 0 eV. The distinct band features observed indicate the metallic nature of \CS, with the Fermi level crossing several bands. This suggests that there are available electronic states at the Fermi level, contributing to the electrical conductivity of the material. The DOS plot provides a detailed view of the contributions from different atomic orbitals. The total DOS is shown alongside the partial DOS for Cr and Si atoms. The overall DOS shows significant contributions near the Fermi level, reinforcing the metallic nature of \CS. The partial DOS for Cr atoms indicates dominant $d$-orbital contributions near the Fermi level. This is expected as Cr $d$-electrons play a crucial role in the bonding and electronic properties of the material. The partial DOS for Si atoms shows contributions primarily from $p$-orbitals, with some hybridization with Cr $d$-orbitals. This hybridization is indicative of the strong covalent interactions between Cr and Si atoms in the \CS~structure. 
The calculated bands and DOS are in good agreement with previous calculations \cite{Ren_2018,wei2018,PAN2021110024}.

The mechanical properties of a material are fundamentally characterized by its elastic constants, including \( C_{11} \) (longitudinal stiffness), \( C_{12} \) (lateral stiffness), and \( C_{44} \) (twisting modulus). The elastic constant \( C_{11} \) quantifies the material's resistance to uniaxial stress applied along the principal crystallographic axes. In contrast, \( C_{12} \) describes the material's response to stress in one direction that induces strain in a perpendicular direction. The elastic constant \( C_{44} \) is crucial for understanding the material's resistance to shear deformation, thereby providing insights into its rigidity and ductility. 
Calculated elastic constants of \CS~are presented in Table \ref{T1} along with elastic constants from previous calculations and experiment. The mechanical stability criteria for a cubic structure are $C_\mathrm{11} >$~0, $C_\mathrm{44} >$~0, and $C_\mathrm{11}~-~C_\mathrm{12}>$~0. These criteria are all satisfied for pristine \CS. The discrepancies between theoretical and experimental values are typical and can be attributed to various factors. In experimental samples, imperfections such as vacancies, dislocations, and grain boundaries are present, which are not accounted for in idealized theoretical models. Additionally, the presence of thermal vibrations at the experimental measurement temperature can lead to reduced elastic constants compared to 0 K calculations. The higher elastic constants in \CS~also suggest that it could be more suitable for high-stress applications compared to other Cr-Si compounds, which may have lower stiffness and potentially higher ductility.
Among the $X_{3}$Si ($X$ = V, Nb, Cr, Mo and W) compounds analyzed~\cite{wei2018stability}, \CS~demonstrates the highest shear modulus (158.9 GPa) and bulk modulus (248.7 GPa), outperforming V$_{3}$Si, Nb$_{3}$Si, and Mo$_{3}$Si, which exhibit lower values across these parameters~\cite{wei2018stability}. Notably, while Mo$_{3}$Si shows a comparable bulk modulus (249.2 GPa), it presents a reduced shear modulus (134.6 GPa), suggesting that \CS~offers superior stiffness and enhanced resistance to shear deformation. These mechanical properties render \CS~the hardest compound in the series, with a hardness of 10.96 GPa. Consequently, \CS~emerges as one of the most mechanically robust $X_{3}$Si compounds, making it particularly suitable for high-stress applications. In contrast, Nb$_{3}$Si, with a lower bulk modulus (175.0 GPa) and hardness (5.72 GPa), may be more ductile but lacks the stiffness required for such demanding environments.
%%%%%%%%%%%%%%%%%%%%%%%%%%%%%%%%%%%%%%%%%%%%%%%%%%%%%%%%%%%%%%%%%%%%%%%%%%%%%%%%%%%%%%%%%%%%%%%%%%%%%%%%%%%%%%%%%%%%%%%
\subsection{Electronic properties of doped-\CS}
Following the examination of the mechanical properties of pristine \CS, we explore potential enhancements through doping with Al, Ga, and Ge. This section focuses on the DOS of the doped systems, with particular attention to the modifications in electronic properties, especially in the vicinity of the Fermi level.
Fig. \ref{Fig9} shows  DOS of Doped \CS~in two different alloy concentrations of 25 and 50 at $\%$. For both 25 and 50 Al doping concentrations, the DOS maintains a metallic character with significant states at the Fermi level. The Cr $d$-orbitals continue to dominate near the Fermi level, similar to pristine \CS. This indicates that the primary electronic structure is retained even with Al doping. The Al $s$ and $p$ orbitals contribute to the DOS but do not significantly alter the overall distribution near the Fermi level, suggesting that Al’s influence on the electronic structure is additive rather than transformative. We see similar behavior for both Ga and Ge, where the metallic character of the structure is maintained in the alloys, with small contributions from the dopants' orbitals

%%%%%%%%%%%%%%%%%%%%%%%
\begin{figure*}
\includegraphics*[scale=0.68]{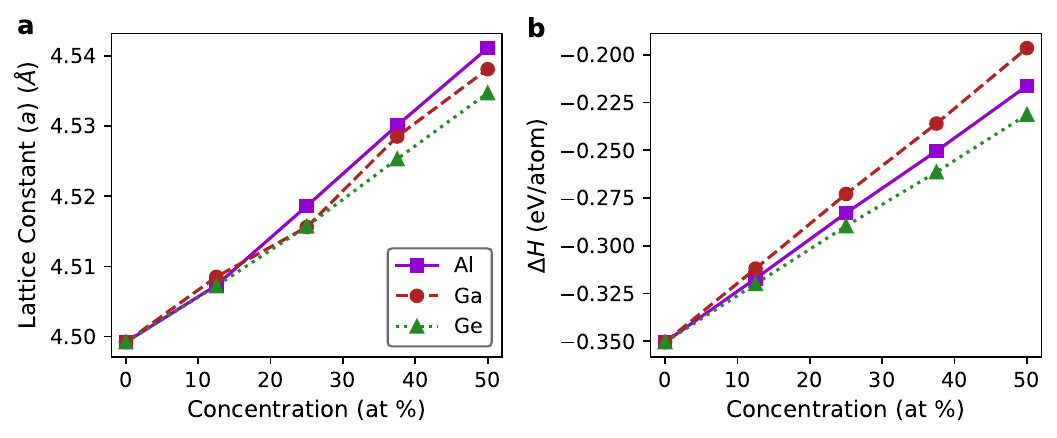}
\caption{
(a) Lattice constants and (b) formation enthalpies of doped \CS~as a function of alloy concentration. 
}
\label{Fig3}
\end{figure*}
%%%%%%%%%%%%%%%%%%%%%%%%%%%%%%5
%%%%%%%%%%%%%%%%%%%%%%%%%%%%%%%%%%%%%%%%%%%%%%%%%%%
\begin{figure*}
\includegraphics*[scale=0.7]{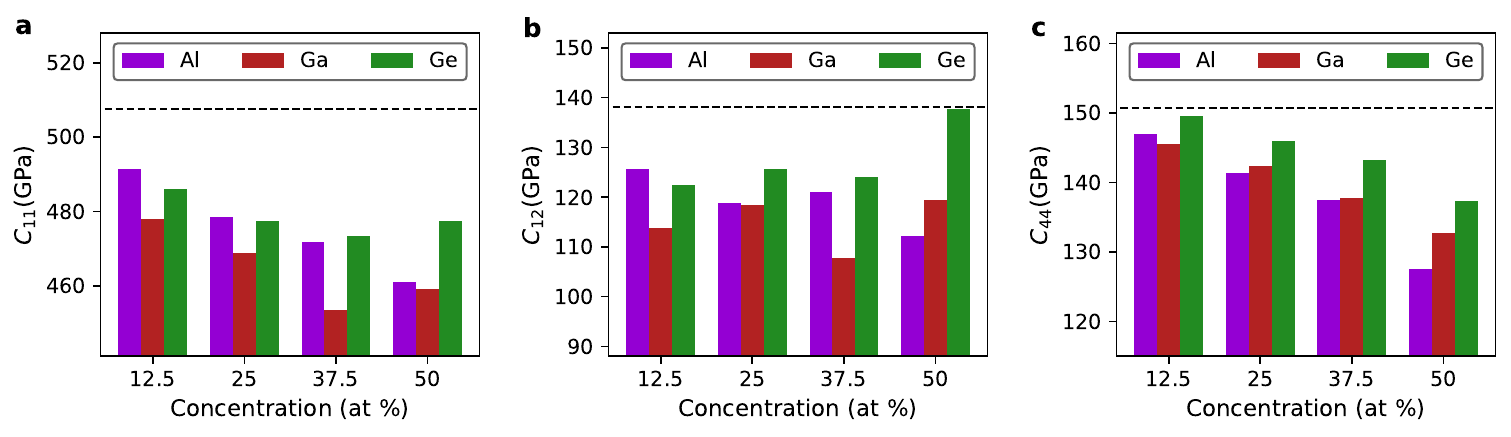}
\caption{
Elastic constants of \CS~doped with Ge, Ga, and Al as a function of alloy concentration. The dotted line shows the calculated elastic constant for the pristine structure. 
}
\label{Fig3_1}
\end{figure*}
%%%%%%%%%%%%%%%%%%%%%%%%%
\begin{figure*}
\includegraphics*[scale=0.68]{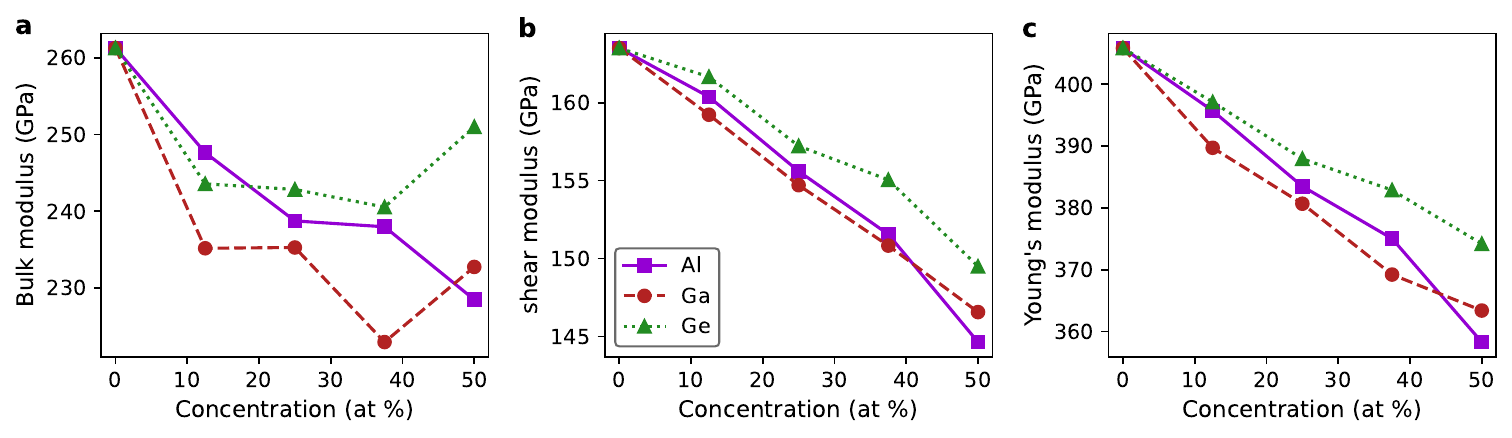}
\caption{
Elastic moduli of doped \CS~as a function of alloy concentration: (a) bulk modulus, (b) shear modulus, and (c) Young's modulus. 
}
\label{Fig4}
\end{figure*}
%%%%%%%%%%%%%%%%%%%%%%%%%%%%%%%%%%%%%%%%%%%%%%%%%%%%%%%%%
\begin{figure*}
	\includegraphics*[scale=0.68]{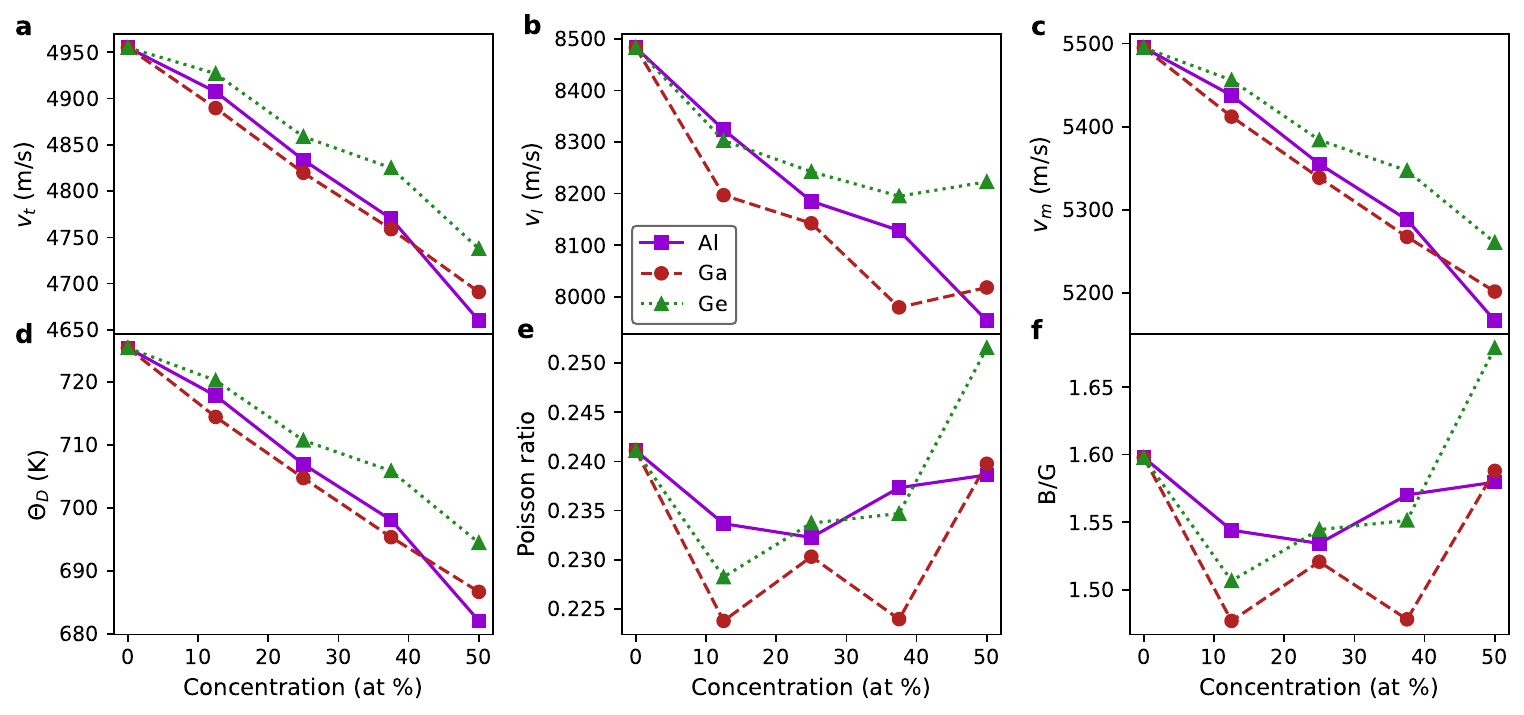}
	\caption{
		(a) Transverse wave velocity, (b) longitudinal wave velocity, (c) average wave velocity, (d) Debye temperature, (e) Poisson ratio, and (f) $B$/$G$ ratio of doped \CS~as a function of alloy concentration. 
	}
	\label{Fig5}
\end{figure*}
%%%%%%%%%%%%%%%%%%%%%%%5
\begin{figure*}
\includegraphics*[scale=0.67]{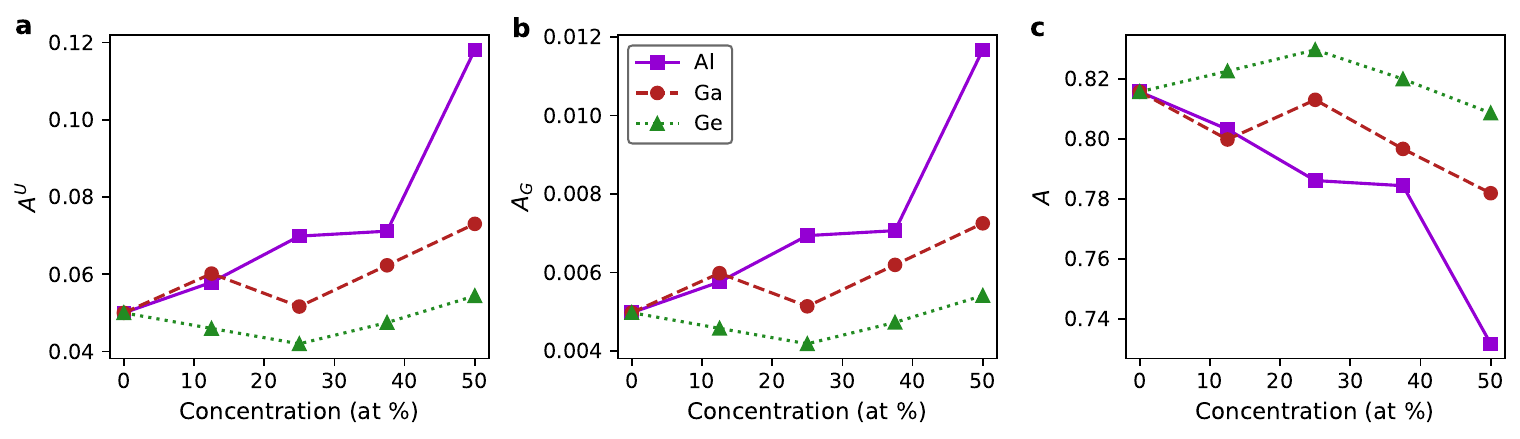}
\caption{
Mechanical anisotropic indices of doped \CS: (a) universal anisotropic index, (b) percent anisotropy of shear modulus, and (c) shear anisotropic factor.
}
\label{Fig6}
\end{figure*}
%%%%%%%%%%%%%%%%%%%%%%%%%%%%%%%%%%%%%%%%%%%%%%%%%%%%%%%%%%%%%%%%%%%%%%%%%%%%%%%%%%%
\begin{figure*}
\includegraphics*[scale=0.36]{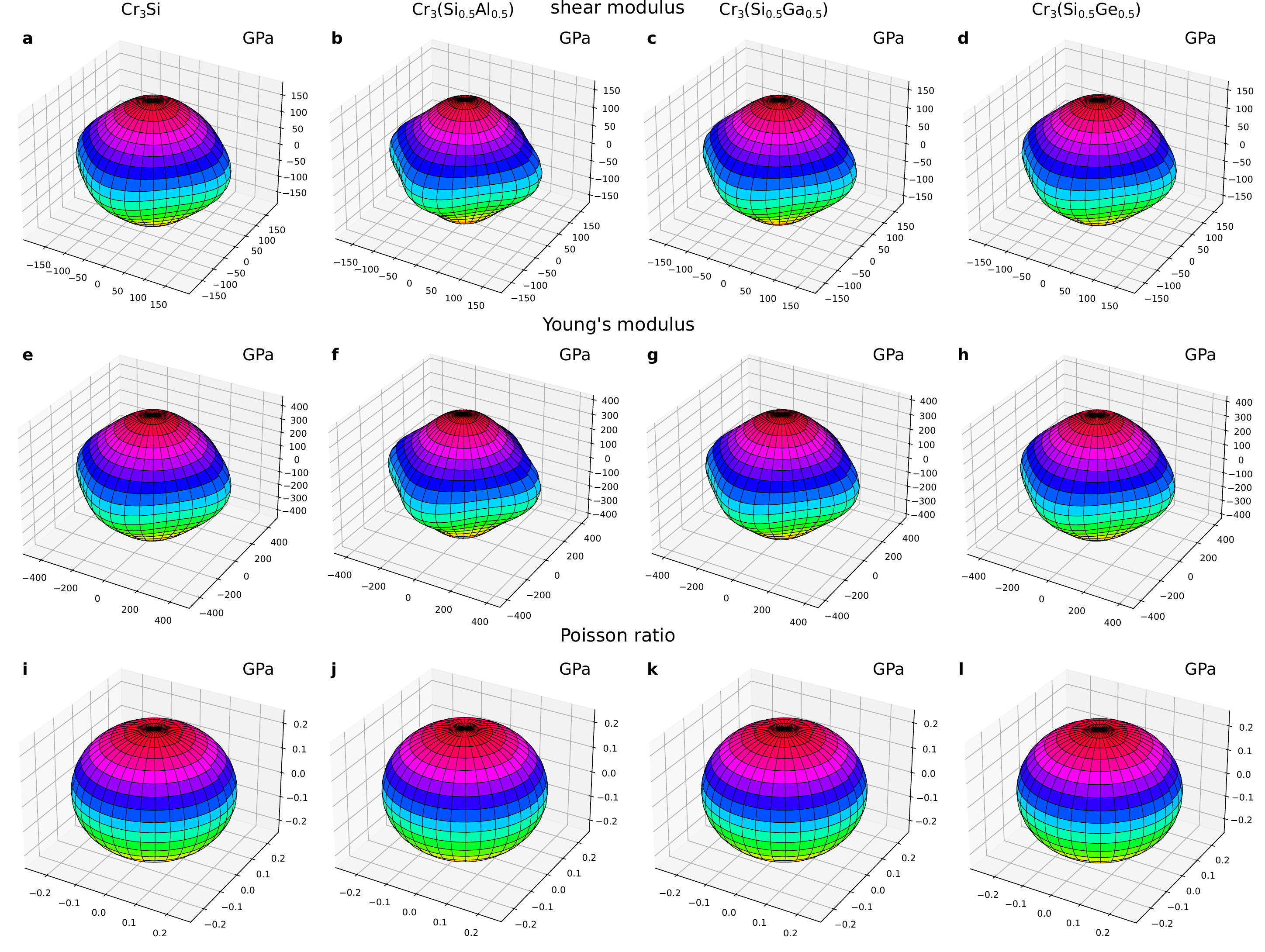}
\caption{
Three dimensional contours of (a)-(d) the shear modulus, (e)-(h) Young's modulus, and (i)-(l) Poisson ratio of doped \CS. The distance between zero and any point on the surfaces is equal to the elastic modulus or Poisson ratio in that direction. 
}
\label{Fig7}
\end{figure*}
%%%%%%%%%%%%%%%%%%
\begin{figure*}
\includegraphics*[scale=0.55]{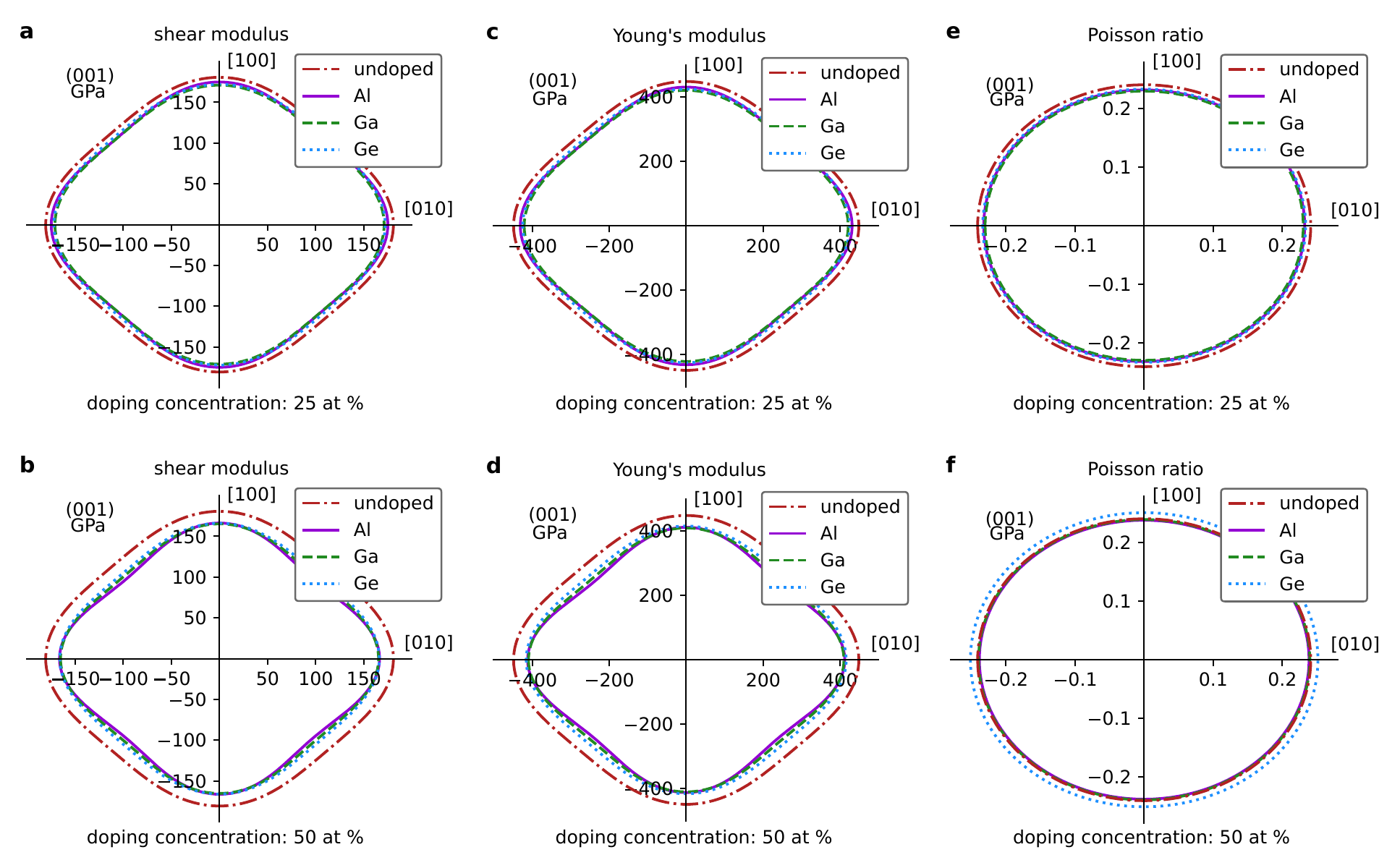}
\caption{
Two dimensional projections of (a),(b) the shear modulus, (c),(d) Young's modulus, and (e),(f) Poisson ratio of doped \CS. 
}
\label{Fig8}
\end{figure*}
%%%%%%%%%%%%%%%%%%%%%%%%%%%%%%%%%%%%%%%%%
Retainment of metallic character indicates that doping with these elements does not disrupt the fundamental electronic structure responsible for metallic conductivity.
The dominance of Cr $d$-orbital contributions near the Fermi level remains consistent from the pristine state (Fig. \ref{Fig2}) through the various doped states. This suggests that the core electronic properties of the \CS~matrix are preserved despite the incorporation of dopants. The similarities between the DOS of doped and pristine \CS~in primary peaks and hybridization patterns indicate that these dopants integrate into the lattice without drastically altering the electronic interactions that define \CS's properteis.
%%%%%%%%%%%%%%%%%%%%%%5
\subsection{Mechanical properties of doped-\CS}
To manipulate the mechanical properties of pristine \CS, one effective approach is to introduce relevant dopants such as Al, Ge, and Ga into the system. In this section, we explore how these dopants influence the mechanical properties of \CS~and discuss the resulting changes in material behavior. In this regard, the first step is to check how the system expands or reduces the lattice constants with doping.
Fig. \ref{Fig3} (a) illustrates the variation in lattice constants of \CS~as a function of alloy concentration for different dopants (Al, Ga, Ge). The lattice constants are shown to increase uniformly with the addition of each dopant. The observed increase in lattice constants with dopant concentration reflects the accommodation of larger dopant atoms within the \CS~structure, leading to lattice expansion. This structural change can impact the mechanical properties of the material, such as its ductility and hardness. 

It is crucial to investigate whether the system is stable after doping or not.
Fig. \ref{Fig3} (b) presents the formation enthalpies of \CS~alloys as a function of dopant concentration. Formation enthalpy is a critical measure of the thermodynamic stability of the alloys. 
Formation enthalpy of undoped \CS~is $-$0.35 eV/atom, in agreement with the experimental value of $-$0.36 eV/atom \cite{CUI2017887}. The consistent negative formation enthalpies across all doped configurations suggest that these alloys are stable, although the degree of stability decreases with higher dopant concentrations. This information is crucial for understanding the trade-offs between structural stability and the desired modifications in material properties through doping.

We now discuss how doping can tune the mechanical properties of cubic \CS. A comparison of the elastic constants of doped \CS~is given in Fig. \ref{Fig3_1}. The graphs show the impact of doping on the stiffness and shear resistance of \CS, with the trends indicating how the introduction of these elements alters the material's mechanical properties. The consistent decrease in $C_{11}$ (longitudinal stiffness), $C_{12}$ (lateral stiffness) and $C_{44}$ (twisting modulus) across all doped configurations suggests that doping generally reduces the mechanical stiffness and shear resistance of \CS. This trend can be attributed to the larger atomic radii of the dopants (Ge, Ga, and Al) compared to Si, leading to lattice expansion and a subsequent weakening of the material's bonding interactions. The most significant reductions are observed with Ga doping, making it the most effective at increasing ductility but also the most detrimental to the material's stiffness and shear resistance. Through these elastic constants, we can find the elastic moduli and subsequently the wave velocities, Debye temperature, and anisotropic indices. This incorporation is explicitly formulated in Appendix \ref{A1}.

The deformation of the material under force can be further examined by looking at the elastic moduli. Fig. \ref{Fig4} presents the elastic moduli (bulk modulus $B$, shear modulus $G$, and Young’s modulus $E$) of \CS~doped with Al, Ga, and Ge at various concentrations. These moduli provide insights into the mechanical properties and the stiffness of the material.
Bulk modulus decreases almost linearly with increasing the Al concentration, indicating a reduction in the material's resistance to uniform compression. This suggests that \CS~becomes more compressible and less stiff as more Al is incorporated. 
The unusual behavior of the bulk modulus for Ga doping, particularly the non-linear fluctuations at higher concentrations, can be attributed to the significant atomic size mismatch between Ga and Si. Ga, having a much larger atomic radius than Si, introduces considerable lattice distortions when incorporated into the \CS~lattice. These distortions lead to non-uniform changes in bonding strength and structural integrity, especially at higher doping concentrations, where the local environment becomes increasingly strained. This strain could result in localized variations in bond stiffness, which manifest as fluctuations in the bulk modulus. Additionally, the lower electronegativity of Ga compared to Si affects the bonding characteristics, further contributing to this behavior. The combination of these factors likely leads to the observed irregularities in the bulk modulus at higher Ga concentrations.
The bulk modulus of Ge-doped \CS~also decreases, but the trend is less steep compared to Al. This implies that Ge doping results in a slightly more stable structure in terms of compressibility compared to Al and Ga.
For all three dopants, the shear modulus decreases with increasing concentration. Shear modulus is crucial for understanding the material’s response to shear stress. Both Al and Ga show significant decrease in shear modulus, indicating that doping with these elements reduces the material's rigidity and makes it more susceptible to shear deformation. Ge also reduces shear modulus, but the reduction is less pronounced compared to Al and Ga. This suggests that Ge-doped \CS~retains more of its original rigidity.
The Young’s modulus, which measures the stiffness of a material under uniaxial stress, decreases steadily with increasing Al concentration. This reduction reflects a decrease in the material's overall stiffness. Similar to shear modulus, Ga-doped \CS~exhibits a decline in the Young’s modulus with increasing concentration, highlighting the reduced stiffness. The decrease in the Young’s modulus for Ge-doped \CS~is more gradual, indicating that the material retains a higher degree of stiffness compared to Al and Ga doping.

The consistent decrease in all three elastic moduli with increasing dopant concentration across Al, Ga, and Ge indicates that doping generally reduces the mechanical stiffness and rigidity of \CS. This trend can be attributed to the larger atomic radii of the dopants, which cause lattice expansion and disrupt the original bonding interactions within the \CS~matrix. The reduction in mechanical strength suggests that heavily doped \CS~may be less suitable for applications requiring high stiffness, and resistance to deformation. However, these materials might still be valuable in applications where reduced stiffness is advantageous or where other properties (such as thermal stability or electrical conductivity) are more critical. By carefully controlling the type and concentration of dopants, it may be possible to tailor the mechanical properties of \CS~for specific applications. For instance, moderate doping levels could achieve a balance between maintaining sufficient mechanical strength while enhancing other desirable properties.

Fig. \ref{Fig5} shows the transverse wave velocity ($v_t$), longitudinal wave velocity ($v_l$), average wave velocity ($v_m$), and Debye temperature  for \CS~doped with Al, Ga, and Ge at various concentrations. These properties provide insights into the mechanical and thermal behavior of the doped materials. The transverse wave velocity, which relates to shear deformation, consistently decreases with increasing dopant concentration for all three dopants. The longitudinal wave velocity, associated with compressional waves, shows a decrease with increased dopant concentration for all dopants, but the trend is less uniform compared to $v_t$. The average wave velocity, a composite measure of $v_t$ and $v_l$, also decreases with higher dopant concentrations for all dopants. The Debye temperature, which correlates with the material's stiffness and thermal properties, follows the same downward trend with increasing dopant concentration.
The consistent reduction in wave velocities and Debye temperature with higher dopant concentrations indicates that doping generally softens the \CS, making it less rigid and more prone to deformation. This is aligned with the trends observed in the elastic moduli (Fig. \ref{Fig4}), further reinforcing the impact of doping on mechanical strength. The decrease in Debye temperature suggests that the thermal conductivity of doped \CS~alloys is reduced. This can be attributed to the larger atomic sizes of the dopants, which disrupt the phonon propagation and reduce thermal conductivity.
While the reduced stiffness might limit structural applications, the altered thermal properties could make doped \CS~suitable for thermal barrier coatings or other high-temperature applications.

Brittleness or ductility of an alloy can be determined using $B$/$G$ and Poisson ($\nu$) ratios \cite{doi:10.1080/14786440808520496}. If $B$/$G$ is higher than 1.75 and $\nu$ is higher than 0.26, the material under investigation is ductile and has strong metallic character, while lower values than the reference points mentioned suggest brittleness. Fig. \ref{Fig5} (e) and (f) show $\nu$ and $B$/$G$ as a function of alloy concentration. The trends for both parameters are quite similar, with the numbers suggesting brittleness of doped \CS~all throughout different percentages. 
Ga and Ge show more pronounced downward behavior in small concentrations compared to Al, suggesting that Ga and Ge doping induce greater brittleness in \CS. The B/G ratio provides similar insights. In our results, the B/G ratio decreases as the doping concentration increases for all dopants, with Al-doped \CS~remaining closer to the ductility threshold, while Ge and Ga-doped \CS~exhibit significantly lower B/G ratios, suggesting a transition towards brittleness at higher doping levels. The trends in both figures reflect the impact of larger atomic radii and lower electronegativities of Ge and Ga compared to Si, which lead to weaker bonds and greater anisotropy, contributing to a reduction in ductility.

Mechanical anisotropic indices provide insight into elastic anisotropy, which is key for understanding the material's behavior under different directional stresses. Fig. \ref{Fig6} visualizes these indices for \CS~doped with Al, Ga, and Ge at various concentrations. 
The universal anisotropic index ($A^U$) quantifies the overall elastic anisotropy of the material, while the percent anisotropy of the shear modulus ($A_G$) measures the directional dependence of shear stiffness. Higher values of $A^U$ and $A_G$ indicate greater anisotropy of the structure. The shear anisotropic factor ($A$) provides a measure of how the shear modulus varies with crystallographic direction. A value of $A = 1$ indicates isotropy, while values deviating from 1 signify anisotropy. 

The increase in anisotropic indices ($A^U$, $A_G$, and $A$) with higher concentrations of Al and Ga suggests that these dopants introduce significant directional dependence in the mechanical properties of \CS. 
This can be traced back to several atomic-level factors, including differences in atomic radii, electronegativity, and
bonding characteristics when compared to Si. Both Al and Ga have larger atomic radii than Si, which leads to lattice distortions when they substitute for Si in the \CS~lattice. These distortions manifest in directional changes in the bonding environment, as larger atoms create asymmetry in the crystal structure, weakening bonds in certain crystallographic directions. This contributes to the observed increase in anisotropy, particularly in the case of Ga, which has a much larger atomic radius than Si. Furthermore, the electronegativity of both Al and Ga is lower than that of Si, meaning that they attract electrons with less strength. This difference in electronegativity alters the nature of the bonds between the dopant atoms and the surrounding Cr atoms. Specifically, the introduction of Al or Ga weakens the overall bonding strength in certain directions, increasing the tendency for the material to deform anisotropically under stress. This is more pronounced with Ga, as the electronegativity of Ga (1.81) is lower than that of Al (1.61), leading to a greater reduction in bond strength and thus a higher degree of anisotropy. Additionally, the bonding characteristics between Cr and the dopant atoms also contribute to the anisotropy. In \CS, the strong covalent bonding between Cr and Si plays a crucial role in maintaining isotropic mechanical properties. When Al or Ga replaces Si, the bond covalency decreases, introducing more metallic character into the bonding network. This metallic nature allows for easier deformation along certain crystallographic axes, further enhancing the anisotropy of the material. Thus, the observed increase in anisotropy with Al
and Ga doping is driven by a combination of larger atomic radii, lower electronegativity, and altered bonding characteristics, all of which weaken the overall structural integrity in specific directions and contribute to anisotropic mechanical behavior. 

This anisotropy can influence the material’s response to mechanical stresses, potentially leading to uneven deformation or failure along specific crystallographic directions. Materials with high anisotropy might be less suitable for structural applications where isotropic properties are critical for uniform load distribution.
Applications that can benefit from directional mechanical properties might leverage higher anisotropy, while those requiring uniform strength might limit doping concentrations.
Fig. \ref{Fig7} shows the three dimensional surfaces of $G$, $E$, and $\nu$ for pristine and doped \CS. Dopants Ge and Ga generate slight changes in the crystal's anisotropy, as we predicted from the anisotropic indices of the alloys. Al creates the most significant changes in the material's anisotropy among the three studied elements when going to higher percentages of doping. This is sufficiently visible when looking at the three dimensional contours of the elastic moduli. To see these changes more clearly, the two dimensional projections of $G$, $E$, and $\nu$ are given in Fig. \ref{Fig8} on the (001) plane at 25 \% and 50 \% alloy concentrations. Besides the alterations in anisotropy by Al, we can also see distinctly that Ge raises $\nu$ at high doping percentages. 

\section{CONCLUSION}\label{IV}

This study comprehensively investigated the effects of Ge, Ga, and Al doping on the mechanical properties and anisotropy of cubic \CS~using first-principles calculations. Our findings reveal that doping with these elements leads to an increase in lattice constants, indicative of lattice expansion due to the larger atomic radii of the dopants. Despite the negative formation enthalpies observed across all doped configurations, indicating overall thermodynamic stability, the mechanical stiffness and rigidity of \CS~decrease with increasing dopant concentration. This is evidenced by the reduction in bulk modulus, shear modulus, and Young’s modulus across the doped samples, suggesting that the material becomes softer and less resistant to deformation as doping concentration increases.
This study further highlights the impact of doping on the mechanical anisotropy of \CS. Al and Ga doping significantly increase the anisotropy indices, suggesting that these dopants introduce considerable directional dependence in the mechanical properties. This enhanced anisotropy could lead to uneven deformation under stress, particularly in applications requiring uniform mechanical strength. In contrast, Ge doping maintains a relatively isotropic behavior, making it a more favorable candidate for applications that demand balanced mechanical and thermal properties.
The wave velocities and Debye temperature analyses further support the conclusion that doping generally softens the \CS~structure, with potential implications for its thermal conductivity. The consistent retention of metallic character across all doped configurations, as indicated by the DOS analyses, suggests that the core electronic properties of \CS~are preserved despite the introduction of dopants. This retention is crucial for applications where electrical conductivity is essential.

In summary, while Ge, Ga, and Al doping can be employed to tailor specific properties of \CS, such as thermal stability and oxidation resistance, these modifications generally come at the cost of reduced mechanical strength and increased anisotropy. 
Ge doping, in particular, presents an ideal balance of maintaining isotropic mechanical properties while introducing beneficial effects such as thermal stability, making it suitable for applications requiring both mechanical robustness and thermal management. Meanwhile, Ga and Al doping, which introduce greater anisotropy, could be explored in scenarios where directional mechanical properties are advantageous, such as in coatings or materials subject to anisotropic stress conditions.
%Ge emerges as the most balanced dopant, offering a trade-off between maintaining structural integrity and enhancing other functional properties. 
These insights provide a valuable foundation for the future design and optimization of \CS-based materials, particularly for high-performance applications where directional mechanical properties and thermal stability are of paramount importance.

\section{ACKNOWLEDGMENTS}
We gratefully acknowledge fruitful discussions with S. Mu. Use was made of computational facilities purchased with funds from the National Science Foundation (CNS-1725797) and administered by the Center for Scientific Computing (CSC). The CSC is supported by the California NanoSystems Institute and the Materials Research Science and Engineering Center (MRSEC; NSF DMR 2308708) at UC Santa Barbara. 

\section{CONFLICT OF INTEREST}
The authors declare no conflict of interest.

\section{DATA AVAILABILITY STATEMENT}
The data that support the findings of this study are available from the corresponding author upon reasonable request. 

\appendix
\section{Calculation of mechanical properties}\label{A1}

The structural stability of a doped system can be determined through formation enthalpy calculation using the following formula \cite{PhysRevB.64.045208}:
  
\begin{align}
\Delta H_{\mathrm{Cr}_i\mathrm{Si}_j\mathrm{X}_k}=\notag \\
\frac{1}{i+j+k}&(E_{\mathrm{Cr}_i\mathrm{Si}_j\mathrm{X}_k}-i\times
E_\mathrm{Cr}-j\times E_\mathrm{Si}-r\times E_\mathrm{X}),
\end{align}

where $i$, $j$, $k$ are the molar fractions of Cr, Si, and the alloying elements X in the alloy, $E_\mathrm{Cr}$, $E_\mathrm{Si}$, $E_\mathrm{X}$ are the energies of pure elements, and $E_{\mathrm{Cr}_i\mathrm{Si}_j\mathrm{X}_k}$ is the energy of the corresponding alloy.

The bulk and shear moduli of a cubic structure can be evaluated using the Voigt-Reuss-Hill (VRH) approximation. The VRH approximation is the average of the lower bound of Voigt and upper bound of Reuss, providing the methodology to connect elastic constants to mechanical properties of crystalline materials. These moduli are obtained as follows \cite{PhysRevB.48.5844}:
\begin{equation}\label{eq2}
B=B_V=B_R=\frac{C_{11}+2C_{12}}{3},
\end{equation}

\begin{equation}
G_V= \frac{C_{11}-C_{12}+3C_{44}}{5},
\end{equation}

\begin{equation}
G_R= \frac{5(C_{11}-C_{12})C_{44}}{3(C_{11}-C_{12})+4C_{44}},
\end{equation}

\begin{equation}
G= \frac{G_V+G_R}{2}. 
\end{equation}

$C_{ij}$ in the above formulas are the calculated elastic constants of the alloys. $B_{V}$, $B_{R}$, and $B$ are the bulk moduli calculated by Voigt, Reuss, and Voigt-Reuss-Hill approximation, respectively. $G_{V}$, $G_{R}$, and $G$ are the shear moduli calculated by Voigt, Reuss, and Voigt-Reuss-Hill approximation, respectively. The isotropic Young's modulus $E$ can then be calculated as:

\begin{equation}\label{eq6}
E=\frac{9BG}{3B+G}.
\end{equation}

The isotropic Poisson ratio $\nu$ is evaluated through:
\begin{equation}\label{eq7}
\nu=\frac{3B-2G}{2(3B+G)}.
\end{equation}

The wave velocities and the Debye temperature $\Theta_D$ can be determined using the mechanical properties \cite{ANDERSON1963909}:
\begin{equation}
v_t={(\frac{G}{\rho})}^\frac{1}{2},
\end{equation}

\begin{equation}
v_l={(\frac{B+\frac{4}{3}G}{\rho})}^\frac{1}{2},
\end{equation}

\begin{equation}
v_m={[\frac{1}{3}(\frac{1}{v_l^3}+\frac{2}{v_t^3})]}^{-\frac{1}{3}},
\end{equation}

\begin{equation}
\Theta_D={\frac{h}{k_B}[\frac{3n}{4\pi}(\frac{N_A\rho}{M})]}^{\frac{1}{3}}v_m,
\end{equation}
where $v_t$, $v_l$, and $v_m$ are transverse, longitudinal, and average wave velocities; $n$ is the number of atoms per formula, $M$ is the molecular weight, $\rho$ is the density and $h$, $k_B$, and $N_A$ are the Avagadro's, Planck's, and Boltzmann constants, respectively.

Elastic anisotropy of a material can be characterized through several anisotropic indices \cite{zener1965elasticity,Vahldiek1968,PhysRevLett.101.055504}:
\begin{equation}
A^U=\frac{5G_V}{G_R}+\frac{B_V}{B_R}-6,
\end{equation}

\begin{equation}
A_B=\frac{B_V-B_R}{B_V+B_R},
\end{equation}

\begin{equation}
A_G=\frac{G_V-G_R}{G_V+G_R},
\end{equation}

\begin{equation}
A=\frac{2C_{44}}{C_{11}-C_{12}}.
\end{equation}

Here $A^U$ is the universal anisotropic index, $A_B$ and $A_G$ are the percent anisotropy of the bulk and shear modulus, and $A$ is the shear anisotropic factor. Elastic anisotropy can provide needed insight for the applications of doped \CS~as it connects directly to the production of micro cracks inside the material. The formation and propagation of these cracks influence the usage of the doped compounds as structural materials at elevated temperatures. 

Elastic anisotropy can be further investigated as a function of crystallographic orientation using the three dimensional surface of an elastic modulus. $E$ of a crystal with cubic structure is formulated as \cite{Nye1985}:
\begin{equation}
\frac{1}{E}=S_{11}-2(S_{11}-S_{12}-\frac{1}{2}S_{44})(l_1^2l_2^2+l_1^2l_3^2+l_2^2l_3^2),
\end{equation}
where $S_{ij}$ are the elastic compliance constants; and $l_1$, $l_2$, and $l_3$ are the direction cosines. We can also calculate $\nu$ and $G$ in different directions using spherical coordinates. This can be done by combining this equation and Eqs. \ref{eq6} and \ref{eq7}.

\bibliography{Cr3Si}
\end{document}